\documentclass[preprint,showpacs,preprintnumbers,amsmath,amssymb]{revtex4}

\usepackage{graphicx,epsf,epsfig}
\usepackage{dcolumn}
\usepackage{bm}
\usepackage{subfigure}

\begin{document}


\title{A Generic Model for Current Collapse in Spin Blockaded Transport}
\author{Bhaskaran Muralidharan and Supriyo Datta}
\affiliation{School of Electrical and Computer Engineering and Network for Computational Nanotechnology, Purdue University, West Lafayette IN-47907, USA}


\date{\today}
\begin{abstract}
A decrease in current with increasing voltage, often referred to as negative differential resistance (NDR), 
has been observed in many electronic devices and can usually be understood within a one-electron picture. 
However, NDR has recently been reported in nanoscale devices with large single-electron charging energies 
which require a many-electron picture in Fock space. This paper presents a generic model 
in this transport regime leading to a simple criterion for the conditions required to observe NDR and 
shows that this model describes the recent observation of multiple NDR's in Spin Blockaded transport 
through weakly coupled-double quantum dots quite well. 
This model shows clearly how a delicate interplay of orbital energy offset, delocalization and Coulomb interaction lead to the observed NDR under the right conditions, and also 
aids in obtaining a good match with experimentally observed features. We believe the basic model could be useful in understanding other experiments in this transport regime as well.
\end{abstract}
\maketitle
\section{Introduction}  
The recent observation of current suppression due to ``Pauli spin blockade" in weakly coupled-double quantum dots (DQDs) 
represents an important step towards the realization and manipulation of qubits \cite{tar,tar2,pet,cm,kw}. 
It is believed that the ``Pauli spin blockade" arises from the occupation of a triplet state \cite{tar} that is filled, but is not emptied easily. Our model supports this picture and puts it in the context of what 
we believe is a much more generic model involving a ``blocking" \cite{het} or ``dark" state \cite{dar}.

A decrease in current with increasing voltage, often referred to as negative differential resistance (NDR),
has been observed in many electronic devices such as degenerately doped bulk semiconductors \cite{esaki}, quantum wells \cite{cap} and even nano-structures \cite{hers}, all of which can usually be understood within a one-electron picture.
However, a number of recent experiments \cite{tar,tar2,tour,kiehl,Heer} have reported NDR in nanoscale devices with large
single-electron charging energies which may require a many-electron picture in Fock space.
Although specific models such as coupling to vibronic states \cite{het}, spin selection rules \cite{rom,timm} asymmetric contact couplings \cite{het}, internal charge transfers \cite{kiehl}, and possibly
conformational changes have been put forth to explain some of these observations, we are not aware of any generic models
comparable in clarity and generality to those available for the occurrence of NDR in the one-electron regime.

The objective of this paper is to present such a model for the strongly correlated transport regime, within the sequential 
tunneling approximation \cite{rralph,timm}, 
where current collapse and NDR arise from the system being locked into a ``blocking" many electron state
that can only be filled from one contact but cannot be emptied by the other. Once it is occupied,
this state blocks any further current flow. This concept (also invoked in a past work \cite{dar}) 
can be used to understand a general category of experiments that involve NDRs.
The basic idea behind current collapse is a condition under
which an excited state of a Coulomb Blockaded channel,
normally inaccessible at equilibrium can, under transport conditions,
be occupied. A drop in the current at a higher bias occurs if this excited state is a ``blocking" state charecterized by 
a slow exit rate. In the first part of this paper we hence obtain a simple criterion for NDR in terms of the rates of filling and emptying of states.

In the later part of this paper, we then apply our model to a specific example in detail, 
namely the NDR observed in weakly coupled-double quantum dots (DQD's) \cite{tar,tar2}. Our analysis, directly addresses current collapse in terms of current magnitudes, without invoking non-equilibrium population analysis \cite{fran,plat}.
Thus current magnitudes at various bias voltages can be solely
expressed in terms of the DQD electronic structure parameters and electrode coupling strengths. 
It then becomes very transparent as to how the interdot orbital offset, hopping, onsite and long
range correlations dramatically affect the observed NDR's, and leakage currents.  
In fact, all the non-trivial features in spin blockade I-V's \cite{tar} such as multiple NDR's, bias
asymmetry in current, gateable current collapse \cite{tar2} and finite leakage currents can
all be explained using the basic ideas developed here. We believe that our criterion obtained from a simple model could be useful for a wide class of experiments beyond the DQD structure discussed in this paper.\\

\section{Current Collapse Mechanism}
Consider a generic Coulomb Blockaded system shown schematically in fig.1a 
that is coupled weakly to contacts defined by chemical potentials $\mu_L$ and $\mu_R$ respectively. Such a Coulomb Blockaded 
system is strongly interacting and is described by its states in Fock space. 
A general condition for NDR to occur can be cast in terms of three Fock space states 
$\mid A \rangle$, $\mid B \rangle$, and $\mid C \rangle$ with energies $E_A<E_B<E_C$ respectively,
shown in fig.1b, that represent three accessible states within the bias range of interest. 
Typically, $\mid A \rangle$, $\mid B \rangle$ could be the ground states of
the $N-1$ and $N$ electron systems, while $\mid C \rangle$, the first excited
state of the $N$ electron system. This description is equally valid for similar transitions between $N$ and $N+1$ electron states. Transport of electrons involves single charge removal or addition between states $\mid B \rangle$, $\mid C \rangle$ and $\mid A \rangle$ that differ by an electron, via
addition and removal transition rates $R_{A \leftrightarrow B,C}$. Such an electron exchange is initiated when
reservoir levels are in resonance with single electron transport channels
$\epsilon_1=E_B-E_A$ and $\epsilon_2=E_C-E_A$ respectively. 
The I-V characteristics shown schematically in fig.1c, of this three state system shows two plateaus with current magnitudes $I_{p1}$ and $I_{p2}$ respectively. Current 
Collapse or NDR occurs when $I_{p1}>I_{p2}$. Let us now derive a condition for
NDR to occur in the positive bias I-V characteristics, of this three state system.

\begin{figure}
\centerline{\epsfig{figure=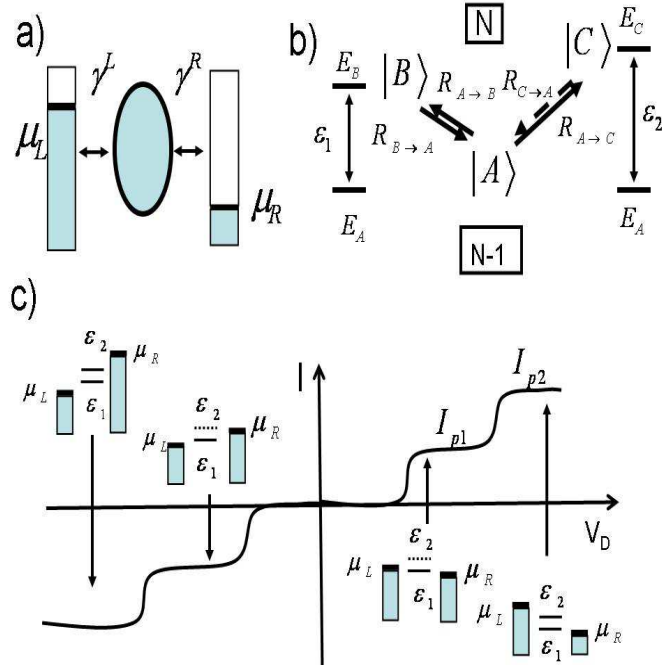,height=3.5in,width=3.5in}}
\caption{a) A generic mechanism for current collapse, rectification and leakage currents through a Coulomb Blockaded system can be 
cast in terms of b) three device Fock-space states, and transitions $\mid A \rangle \leftrightarrow \mid B \rangle$
and $\mid A \rangle \leftrightarrow \mid C \rangle$ between those that differ by a single electron. c) Typical
I-V characteristics of this system comprises of two break points following a sequential
access of two transport channels $\epsilon_1$ and $\epsilon_2$. Insets depict
the bias configurations that corresponds to various plateau currents. The central result of the paper, i.e., 
the condition for current collapse to occur ($I_{p1}>I_{p2}$), under positive or negative bias conditions, 
derived in Eqs.~\ref{eq:NDR} and ~\ref{eq:NDR2}, is dictated by the rate of 
de-populating the ``blocking" state $\mid C \rangle$. Such transitions are represented via dotted arrows in b).} 
\label{fig_1}
\end{figure}

These plateau currents can be
evaluated in terms of bias dependent probabilities of the three states $\mid A \rangle$,
$\mid B \rangle$ and $\mid C \rangle$, given by a set of coupled
master equations \cite{rralph,bru,mat,mitra,koch}:
\begin{eqnarray}
\frac{dP_A}{dt} &=& -\left (R_{A \rightarrow B}+R_{A \rightarrow C} \right ) P_A + R_{B \rightarrow A} P_B + R_{C \rightarrow A} P_C \nonumber\\
\frac{dP_B}{dt} &=& -R_{B \rightarrow A}P_B + R_{A \rightarrow B} P_A \nonumber\\
\frac{dP_C}{dt} &=& -R_{C \rightarrow A} P_C + R_{A\rightarrow C} P_A,
\label{eq:m1}
\end{eqnarray}
where $R_{I \rightarrow J}$ denotes the transition rate between state $\mid I \rangle$ and $\mid J \rangle$,
which could be an addition or a removal transition between these states that differ by
a single electron. These rates are most generally described as
\begin{eqnarray}
R_{A \rightarrow B} &=& \sum_{\alpha=L,R}\gamma^{\alpha} M^{\alpha}_{AB}  f_{\alpha}(\epsilon_1) \nonumber\\
R_{A \rightarrow C} &=& \sum_{\alpha=L,R}\gamma^{\alpha} M^{\alpha}_{AC}  f_{\alpha}(\epsilon_2),
\label{eq:rates}
\end{eqnarray}
where $\gamma^{L,R}$ denotes the coupling strength to either contacts, $M$'s denote coherence factors and 
$f_{L,R}(E)=f(E-\mu_{L,R})$ denote Fermi-Dirac distributions in the leads. The coupling strengths are often
expressed in terms of the tunneling Hamiltonian matrix elements $\tau_{k\alpha,m}$, where $k \alpha$ represents the $k^{th}$ eigen mode
in $\alpha=L,R$ either contact \cite{mwl}, and $m$ denotes a quantum dot state as:
$\gamma^{\alpha}=2\pi \sum_{k} \mid \tau_{k \alpha,m} \mid^2 \delta(E-\epsilon_{k \alpha})$.
Note that for the downward transitions the $f$'s are replaced
by $1-f$'s. The coherence factors depend significantly on the structure of the many-body states 
involved in the transition and will be considered while dealing with specific examples 
in the following sections. We also note that $R_{B \rightarrow C} = R_{C \rightarrow B} = 0$, since transitions
between states with equal electron numbers are forbidden, in the absence of
coupling to radiation fields \cite{het}, or higher order tunneling \cite{rralph}. The steady-state current $I$ equals
both the left and right terminal currents ($I_L$,$I_R$) given by
\begin{eqnarray}
I_L &=& \frac{q^2}{\hbar} (R^{L}_{A \rightarrow B} P_A -R^{L}_{B \rightarrow A} P_B \nonumber\\
\qquad &+& R^{L}_{A \rightarrow C} P_A -R^{L}_{C \rightarrow A} P_C) \nonumber\\
I_R &=& -\frac{q^2}{\hbar} (R^{R}_{A \rightarrow B} P_A - R^{R}_{B \rightarrow A} P_B \nonumber\\
\qquad &+& R^{R}_{A \rightarrow C} P_A - R^{R}_{C \rightarrow A} P_C)  ),
\label{eq:curr}
\end{eqnarray}
where $q$ is the electronic charge and $R^{L,R}_{I \leftrightarrow J}$, stands for
the left or right contact contribution to the transition $\mid I \rangle
\leftrightarrow \mid J \rangle$. Positive bias plateau currents, shown schematically in fig.1c, (see appendix I), 
are given by:
\begin{eqnarray}
I_{p1} &=& \frac{q^2}{\hbar}\frac{\gamma^{L}_{AB} \gamma^{R}_{BA}}{\gamma^{L}_{AB} + \gamma^{R}_{BA}} \nonumber\\
I_{p2} &=& \frac{q^2}{\hbar} \frac{\gamma^L_{AB} + \gamma^L_{AC}}{1 + \frac{\gamma^L_{AB}}{\gamma^R_{BA}} + \frac{\gamma^L_{AC}}{\gamma^R_{CA}}}.
\label{eq:curs}
\end{eqnarray}
where
\begin{equation}
\gamma^{\alpha}_{IJ}=\gamma^{\alpha} M^{\alpha}_{IJ}, 
\end{equation}
When $I_{p2}<I_{p1}$, current drops once the transition $\mid A \rangle \leftrightarrow \mid C \rangle$ 
is accessed. Using the above expressions for $I_{p1}$ and $I_{p2}$, this leads to a simple condition:
\begin{equation}
\frac{1}{\gamma^R_{CA}}>\frac{1}{\gamma^L_{AB}}+\frac{1}{\gamma^R_{BA}}, \qquad (\mu_L>\mu_R)
\label{eq:NDR}
\end{equation}
for current collapse or NDR in the I-V characteristics to occur along the positive bias. A similar condition for  
negative bias NDR is obtained by replacing $L$ with $R$ and vice versa:
\begin{equation}
\frac{1}{\gamma^L_{CA}}>\frac{1}{\gamma^R_{AB}}+\frac{1}{\gamma^L_{BA}}, \qquad (\mu_R>\mu_L).
\label{eq:NDR2}
\end{equation}
The positive bias leakage current is now dictated by $\gamma^R_{CA}$, which corresponds to the rate of leakage from the blocking channel
$\mid C \rangle$. For $I_{p1} \geq I_{p2}$,
currents rise just like a regular Coulomb Blockade staircase due to access of a
second transport channel. Asymmetry of this blockade leads to current rectification 
observed in experiments \cite{tar}, is dictated by the ratio:
\begin{equation} 
\Gamma=\frac{I_{p2}(V_D<0)}{I_{p2}(V_D>0)} \neq 1.
\label{eq:NDR3}
\end{equation}
Thus current collapse can occur along both bias directions,
provided Eqs.~\ref{eq:NDR},~\ref{eq:NDR2} are simultaneously satisfied. Finally, 
asymmetric I-V's result when Eq.~\ref{eq:NDR3} is satisfied, as shown in fig. 2b. The condition for asymmetry to occur 
is just based on the ratio between positive and negative bias plateau currents, and generally depends on various couplings. 
\begin{figure}
\centerline{\epsfig{figure=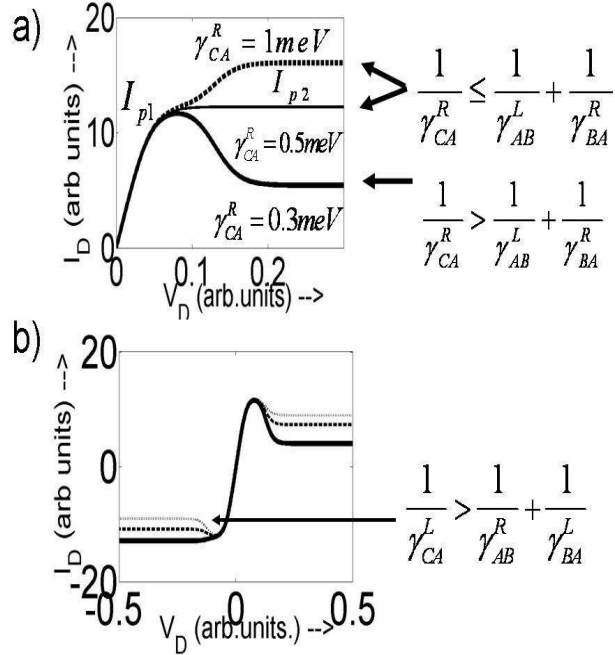,height=3.5in,width=3.5in}}
\caption{I-V characteristics: a) Forward bias I-V characteristics of the three state system with $\gamma^L_{AB}$ and $\gamma^R_{AB}$ 
set to $1$ meV. As $\gamma^R_{CA}$ is varied,
the plateau currents $I_{p1}$ and $I_{p2}$ resulting from a serial access of transport channels $\epsilon_1$ and $\epsilon_2$ with bias, results in a staircase $I_{p1}>I_{p2}$, saturation $I_{p1}=I_{p2}$ or NDR $I_{p2}<I_{p1}$.
Forward bias NDR is achieved when Eq.~\ref{eq:NDR} is satisfied, say when $\gamma^R_{CA}=0.3$meV.
The magnitude of this blockade current $I_{p2}$ is then governed by $\gamma^R_{CA}$.
b) Reverse bias NDR due to the condition (Eq.~\ref{eq:NDR2}), is governed by the removal rate $\gamma^L_{CA}$.}
\label{fig_2}
\end{figure}
Thus, Eq.~\ref{eq:NDR},~\ref{eq:NDR2},and~\ref{eq:NDR3}, summarize the central concept of this paper. 
We will show in the next section that they provide a clear explanation for experimental observations reported for a weakly-coupled DQD system.
It is worth noting that these conditions can be generalized even if the Fock space states $\mid A \rangle$, $\mid B \rangle$ and $\mid C \rangle$ have specific degeneracies \cite{bask_the}
provided the rate constants $\gamma^{L,R}_{ij}$ are modified by appropriate degeneracy factors. At this point it is also worth noting that the degeneracies considered here are such that individual processes
that are involved in the transitions are uncorrelated. This justifies the use 
of rate equations ignoring off-diagonal effects.
Inclusion of off-diagonal terms, via a density matrix rate equation \cite{rbraig} may be required to capture other subtle degeneracies \cite{koenig}
not considered here.

\begin{figure}
\centerline{\epsfig{figure=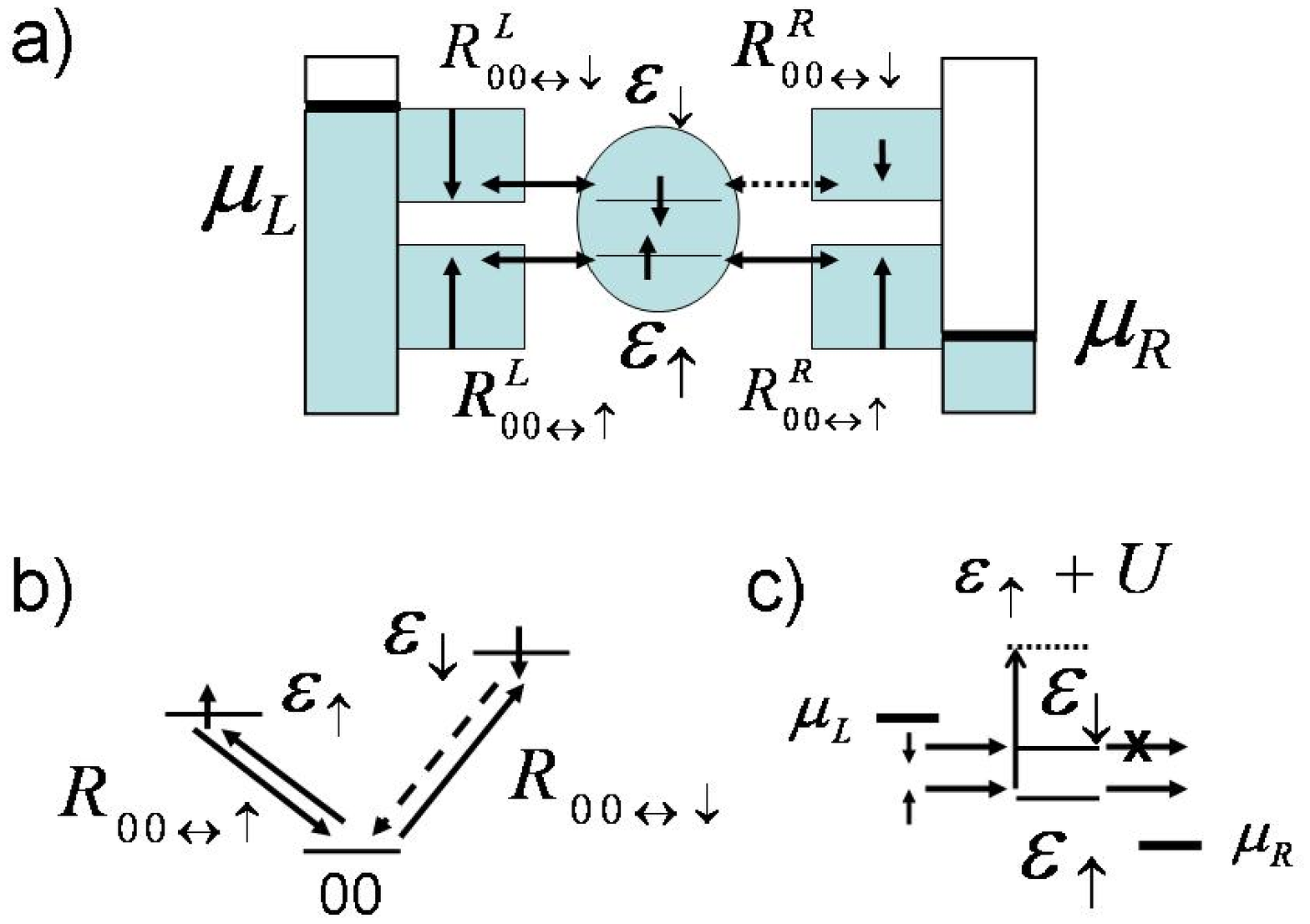,height=3.5in,width=3.5in}}
\caption{a) A simple physical realization of the above NDR condition
using a single interacting quantum dot with spin split levels coupled
to two reservoirs, one (say the left) that is unpolarized and the other (right)
one spin polarized in the up direction. b) This situation automatically satisfies
the NDR condition with $\mid \downarrow \rangle$ acting as the blocking state $\mid C \rangle$,
as shown in the Fock space diagram. c) An equivalent transport energy
diagram simply implies that the conducting state transport channel $\epsilon_{\uparrow}$ is lifted
out of the bias window by an amount $U$ due to interaction with $\epsilon_{\downarrow}$
blocked inside the device, since $\gamma^R_{\downarrow} \approx 0$. $\mu_L >>  \mid \epsilon_{\downarrow} \pm \frac{k_BT}{q} \mid$ ensures that the blocked electron is not reinjected to the left contact. }
\label{fig_3}
\end{figure}

Before, we discuss coupled quantum-dots, it is useful to look into a simpler
example discussed by various works \cite{timm,koenig} and see how it maps onto our generic ``blocking state" model (fig.1b).
Consider a single quantum dot with two non-degenerate levels, one up-spin and one down spin, as shown in fig.3a.
The Fock space picture in fig.3b looks just like
fig. 1b with $\mid A \rangle$ being the state with no electron $\mid 00 \rangle$, $\mid B \rangle$ being the state with one up electron $\mid \uparrow \rangle$, $\mid A \rangle$ being the state with no electron $\mid 00 \rangle$,
and $\mid C \rangle$ being the state with one down-electron $\mid \downarrow \rangle$. Note that the state with two electrons is assumed to have too high
an energy to be accessed in the bias range of interest. It is easy to see that for positive bias that
$\gamma^L_{AB} = \gamma^L_{\uparrow}$, $\gamma^L_{AC} = \gamma^L_{\downarrow}$, 
$\gamma^R_{AB} = \gamma^R_{\uparrow}$, and $\gamma^R_{AC} = \gamma^R_{\downarrow}$,
where $\gamma^{L,R}_{\uparrow, \downarrow}$ denotes the contact couplings with up and down spin reservoir channels. Ordinary contacts typically have
$\gamma_{\uparrow}=\gamma_{\downarrow}$, so that the NDR criteria (Eqs.~\ref{eq:NDR},~\ref{eq:NDR2}) are never satisfied.
However, if one contact, say the right is spin polarized $\gamma^R_{\downarrow}<<\gamma^{R}_{\uparrow}$ then NDR criteria is met for
positive bias though not for negative bias.
\section{Pauli Blockade in Weakly Coupled Double Quantum Dots}
Our idea of blocking states involves an inherent asymmetry within the transport problem.  
In our previous example (fig.3), this was achieved by explicitly postulating a spin polarized contact 
as the asymmetry mechanism. Asymmetries leading to blocking states, 
can also arise from asymmetries within the system's internal degrees, thus requiring no apriori conditions on bare physical contact couplings.
The asymmetry effects now enter the model through coherence factors mentioned earlier in Eq.~\ref{eq:rates}.
The rest of this paper focuses on how these coherence factors could lead
to the NDR conditions Eqs.~\ref{eq:NDR},~\ref{eq:NDR2} derived in the earlier section.

We consider, a DQD system shown in fig.4a, with orbital energies $\epsilon_1
\ne \epsilon_2$, and specifically $\epsilon_1>\epsilon_2$ on each dot. 
We will now show that, for specific orbital offsets $\Delta \epsilon=\mid \epsilon_2-\epsilon_1 \mid$, 
its interplay with hybridization $t$, and Coulomb repulsion $U_{mp}$ can result 
in a current collapse following the notion developed in the past section. 
The I-V characteristics discussed in this section,
closely match the experimental features \cite{tar} that indicate Pauli spin Blockade. 
\begin{figure}
\centerline{\epsfig{figure=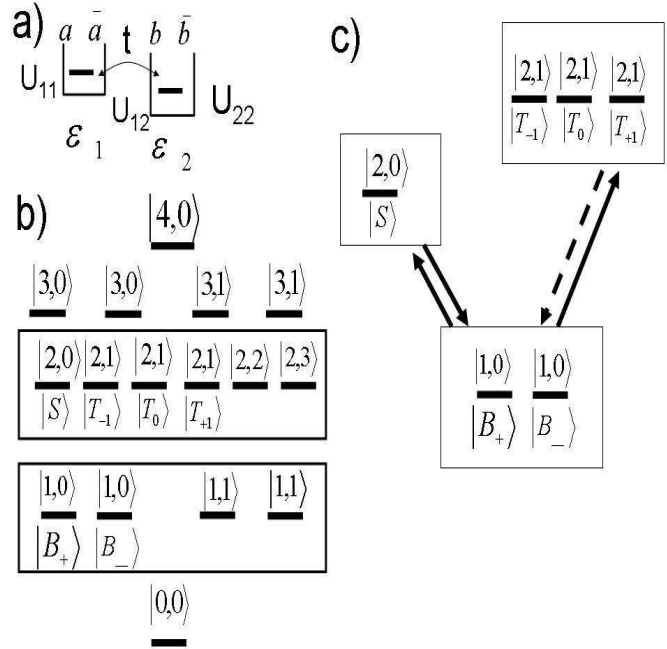,height=3.5in,width=3.5 in}}
\caption{Many electron spectrum of a DQD system. a)
Orbital energies of the two spin degenerate ground levels $a,\bar{a}$ and $b,\bar{b}$ on each dot,
 are given by $\epsilon_1$ and $\epsilon_2$. On-site and long range Charging parameters are given by $U_{11},U_{22}$ and
$U_{12}$ respectively. The interdot hopping parameter is $t$. b) The many-electron spectrum
comprises of $16$ states as shown in Fig.4b, labeled as $\mid N,i \rangle$. States within a charge subspace are just numbered and are not necessarily in any energetic order. The $N=1$ and $N=2$ spectrum are
relevant to the present discussion. The three-state system responsible for spin blockade has degeneracies and
consists of two bonding $\mid B \rangle$, and one singlet $\mid S \rangle$, and three triplet $\mid T \rangle$ respectively. c) The first two transitions involved in the
spin blockade NDR are the two transport channels which can be recast in the familiar tristate form shown (fig. 1b). }
\label{fig_4}
\end{figure}
\subsection{Electronic Structure of a Double Quantum dot}
In this paper, we consider a double quantum dot as shown in Fig.4a, that comprises of
two dots whose lowest spin degenerate orbitals are labeled
$a,\bar{a}$ and $b,\bar{b}$ respectively. Our analysis is limited to 
one orbital per quantum dot thus restricting our analysis to a few meV along either bias directions. 
Higher orbitals on each quantum dot can in a similar way result in multiple NDR's \cite{tar} 
along either bias directions. These higher orbital effects are accessed at a much higher bias than those considered here. 
We believe that the higher bias features are an extension of the general idea developed here, which will be addressed 
in a future work. Let us now define the DQD Hamiltonian \cite{fran,plat2,sand} in a localized 
orbital basis as:
\begin{eqnarray}
H_{QD} &=& \sum_{m} \epsilon_m n_m + \sum_{p} \left ( t_{mp} c^{\dagger}_{m} c_{p} + c.c \right ) \nonumber\\
\qquad &+& U_{m \uparrow m \downarrow} n_{m \uparrow} n_{m \downarrow} + \frac{1}{2} \sum_{p} U_{mp} n_m n_p,
\label{eq:hqd}
\end{eqnarray}
where the left and right dot single particle states are $\mid a \rangle$, $\mid \bar{a} \rangle$ and  $\mid b \rangle$, $\mid
\bar{b} \rangle$ respectively, $\bar{a}$, $\bar{b}$ denoting the opposite spin state. 
$t_{mp}$ denotes the hopping term between the two dots, and 
$U_{m \uparrow m \downarrow}$, and $U_{mp}$ denote the on-site 
and long-range Coulomb repulsion term respectively. 
The many-electron spectrum shown in fig.4b comprises of $16$ states
labeled as $\mid N,i \rangle$ defined by the number of electrons
$N$ and the excitation number $i$. Six states that are relevant to our spin blockade discussion are
labeled within the boxed sections in fig.4b, and form a subset of the $N=1$ and $N=2$ blocks.
The six states consist of a doubly degenerate bonding $\mid 1,0 \rangle$ state in the $N=1$ subspace:
\begin{eqnarray}
\mid B_+ \rangle = \xi \mid a  \rangle + \kappa \mid b \rangle \nonumber\\
\mid B_{-} \rangle = \xi \mid \bar{a}  \rangle + \kappa \mid \bar{b} \rangle ,
\label{eq:ba}
\end{eqnarray}
a singlet $\mid 2,0 \rangle$ state in the $N=2$ subspace:
\begin{eqnarray}
\mid S \rangle &=& \alpha \left ( \mid a \bar{b} \rangle - \mid \bar{a} b \rangle \right ) \nonumber\\
\qquad &+& \beta \mid a \bar{a} \rangle + \delta \mid b \bar{b} \rangle,
\label{eq:singlet}
\end{eqnarray}
and a three-fold degenerate triplet $\mid 2,1 \rangle$ states also in the $N=2$ subspace given by:
\begin{eqnarray}
\mid T_{+1} \rangle &=& \mid a b \rangle \nonumber\\
\mid T_{0} \rangle &=& \frac{1}{\sqrt{2}} ( \mid a \bar{b} \rangle + \mid \bar{a} b \rangle ) \nonumber\\
\mid T_{-1} \rangle &=& \mid \bar{a} \bar{b} \rangle.
\label{eq:triplet}
\end{eqnarray}
Here $\kappa, \xi, \alpha, \beta$, and $\delta$ represent the wavefunction coefficients for
the various basis Fock-states. We shall see in the following subsection that these wavefunction coefficients depend on the physical parameter set $\epsilon_{1},\epsilon_2,t,U_{11},U_{22},U_{12}$ of the DQD, and hence appear in the calculation of
various coherence factors (Eq.~\ref{eq:rates}) involved in the transition rates. It is worth noting 
that the anti-bonding level $\mid 1,1 \rangle$ of the $N=1$ spectrum is not involved in the 
Pauli-Blockade NDR. This is because in the weakly coupled DQD that we consider here, 
the bonding-antibonding gap is directly proportional to the orbital energy offset $\Delta \epsilon$, which 
is of the order of a few milli-volts. The energy scale relevant to our NDR discussion is 
the singlet-triplet splitting which is of a couple of orders of magnitude smaller given that the DQD system is weakly coupled. The anti-bonding state then becomes relevant beyond 
the bias range considered here where the spin-blockade is lifted.  

{\it{Coherence Factors:}} Although six states, (fig.4b) are involved in the spin blockade
I-V characteristics within the bias range of interest, they can be cast in the familiar
tri-state form (fig.1b) as shown in fig.5, by incorporating the appropriate degeneracy 
factors \cite{bask_the} for the transition rates. The degenerate states may just be lumped as one state under such specific conditions \cite{bask_the}.
The coherence factors are evaluated between two states that differ by an electron,
whose structure and symmetry properties depend on the physical parameter set of the DQD and how the individual
dots couple to the two electrodes. The general
definition is as follows:
\begin{equation}
\gamma_{ij}^{\alpha} = \gamma^{\alpha}|\langle N,i|c^\dagger_m|N-1,j\rangle|^2,
\label{eq:tran}
\end{equation}
where $c^\dagger_m$ is the creation/annihilation operators for an electronic state
on the end dot coupled with the corresponding electrode, and $\gamma^{\alpha}$ is the bare left or right electrode coupling factor defined in section I. 
In our case, the coherence factors are evaluated between the a) $\mid 1,0 \rangle$ bonding
and $\mid 2,0 \rangle$ singlet:
\begin{eqnarray}
\gamma^L_{00} &=& \gamma^L \mid \langle S \mid c_{a,\bar{a}}^{\dagger} \mid B_{-} (B_+) \rangle \mid^2 \nonumber\\
&=& \gamma^L (\beta \xi + \alpha \kappa)^2 \nonumber\\
\gamma^R_{00} &=& \gamma^R \mid \langle S \mid c_{b,(\bar{b})}^{\dagger} \mid B_{-} (B_+ ) \rangle \mid^2 \nonumber\\
&=& \gamma^R (\alpha \xi + \kappa \delta)^2,
\label{eq:rate_bs}
\end{eqnarray}
b) $\mid 1,0 \rangle$ bonding and $\mid 2,1 \rangle$ triplet states:
\begin{eqnarray}
\gamma^L_{10} &=& \gamma^L ( \mid \langle T_0 \mid c_{a}^{\dagger} \mid B_{-} \rangle \mid^2 \nonumber\\
&+& \mid \langle T_0 \mid c_{\bar{a}}^{\dagger} \mid B_+ \rangle \mid^2 )\nonumber\\
&=& \mid \langle T_{\pm1} \mid c_{a,\left(\bar{a} \right )}^{\dagger} \mid B_{\mp} \rangle \mid^2  = \gamma^L{\kappa}^2\nonumber\\
\gamma^R_{10} &=& \gamma^R ( \mid \langle T_0 \mid c_{b}^{\dagger} \mid B_{-} \rangle \mid^2 \nonumber\\
&+& \mid \langle T_0 \mid c_{\bar{b}}^{\dagger} \mid B_+ \rangle \mid^2) \nonumber\\
&=& \mid \langle T_{\pm1} \mid c_{b,\left(\bar{b} \right )}^{\dagger} \mid B_{\mp} \rangle \mid^2 = \gamma^R {\xi}^2.
\label{eq:rate_bt}
\end{eqnarray}
Recall that the transport feature central to the Pauli Blockade experiments
is the occurrence of multiple NDR's \cite{tar} along both bias directions, and this condition must be recast into the model developed in the first section.
Therefore, the basic criterion for NDR to occur along the positive bias ($V_D>0$) direction is:
\begin{equation}
\frac{1}{2\gamma^{R}_{10}}> \frac{1}{\gamma^{L}_{00}} + \frac{1}{2\gamma^{R}_{00}},
\label{eq:sbc1}
\end{equation}
and along the negative bias direction ($V_D<0$) is:
\begin{equation}
\frac{1}{3\gamma^{L}_{10}}> \frac{1}{2\gamma^{L}_{00}} + \frac{1}{\gamma^{R}_{00}}, 
\label{eq:sbc2}
\end{equation}
Notice that NDR conditions in the above equations also 
include degeneracy factors as depicted in fig.5. 
 A direct substitution from Eqs~\ref{eq:rate_bs} and ~\ref{eq:rate_bt} with $\gamma^L=\gamma^R$ yields
\begin{equation}
\frac{1}{2 \xi^2}> \frac{1}{ (\beta \xi + \alpha \kappa)^2} + \frac{1}{2(\alpha \xi + \kappa \delta)^2} 
\label{eq:sbc1b}
\end{equation}
along the positive bias direction and
\begin{equation}
\frac{1}{3 \kappa^2}> \frac{1}{2 (\beta \xi + \alpha \kappa)^2} + \frac{1}{ (\alpha \xi + \kappa \delta)^2}. 
\label{eq:sbc2b}
\end{equation}
along the negative bias direction. To apply these relations we need to consider the wavefunction coefficients $\xi$,$\kappa$,$\alpha$,$\beta$, and $\delta$ in more detail.

\subsection{Spin Blockade Transport}
Consider first the wavefunction coefficients $\kappa$ and $\xi$ for the $N=1$ bonding state $\mid B \pm \rangle$. It is easily seen that $\kappa \approx \xi$ for $\Delta \epsilon << t$ and $\kappa >> \xi$ when $\Delta \epsilon >> t$, given that $\epsilon_1 > \epsilon_2$. The two electron triplet 
(see fig. 5) is relatively un-affected by electronic structure variations, but the singlet state described by the wavefunction coefficients $\alpha,\beta$ and $\delta$ requires some more discussion.

The presence of a single electron (fig.6) in one dot influences the addition of another on either of 
them as a result of on-site or long range Coulomb interaction.  For $t << \Delta \epsilon,U$ relative magnitudes of $\alpha$, $\beta$ and $\delta$ now depend on off-sets between the total energies of the three configurations shown in fig.6 which we shall consider now.  
\begin{figure}
\centerline{\epsfig{figure=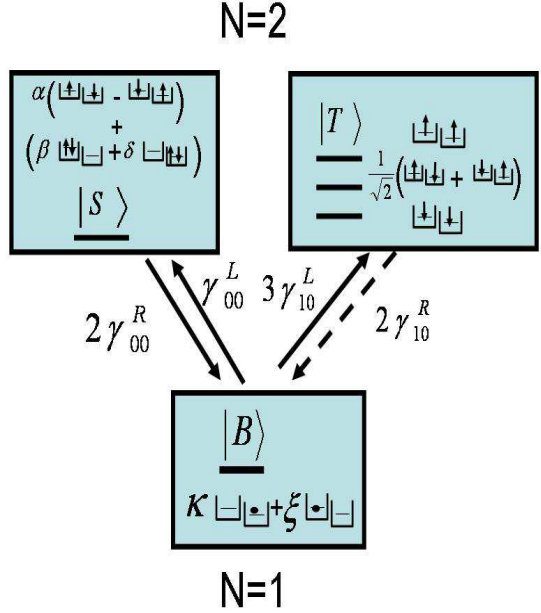,height=3.5in,width=3.5 in}}
\caption{Internal structure of the Three state system: This three-state system
responsible for spin blockade comprises of one-electron Bonding $\mid B \rangle$, and two electron singlet $\mid S \rangle$, and triplet $\mid T \rangle$ respectively.
Electronic configuration of the three states is also shown along side.
The singlet, triplet and bonding states form the trio of Fock space states involved in the SB I-V characteristics,
with the triplet state being the blocking state.}
\label{fig_4}
\end{figure}

{\it{Transport Scenarios:}} 
Consider three transport scenarios depicted in fig.7, which are based 
on the relative offsets between orbital energies $\Delta \epsilon =\epsilon_1 - \epsilon_2$, 
and relative offsets between the two electron wavefunction configurations (fig.6) $\Delta E_{1}=\epsilon_{1}+U_{11}-\epsilon_2-U_{12}$, and $\Delta E_{2}=\epsilon_{2}+U_{22}-\epsilon_1-U_{12}$.
In all our simulations we have used parameters consistent with a weakly-coupled  DQD structure in \cite{tar}, with $t=0.2-0.3$meV, $U_{11}=U_{22}=4$meV, and $\gamma^L=\gamma^R=0.01$meV, while other parameters are varied to the three cases depicted in Fig.7.

\begin{figure}
\centerline{\epsfig{figure=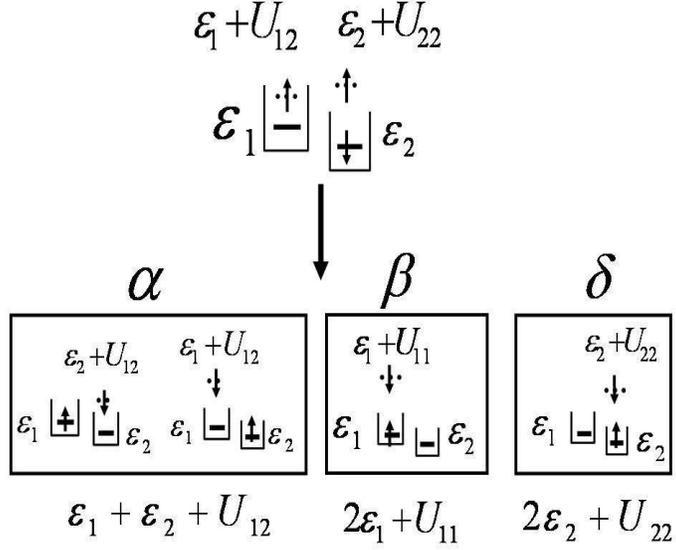,height=3.25in,width=3.6in}}
\caption{Understanding the internal structure of the $2$ electron singlet: 
Here, both on-dot $U_{mm}$ and long range Coulomb repulsion $U_{mp}$ affect the wavefunction,
by raising the onsite energy of the second electron depending on where the first electron is localized. 
The singlet state is a superposition between states with one electron in each dot ($\mid a
\bar{b} \rangle$, $\mid \bar{a} b \rangle$, $\mid a b \rangle$,$\mid \bar{a}
\bar{b} \rangle$), and the ones with both on either dots ($\mid a \bar{a}
\rangle$ and $\mid b \bar{b} \rangle$). A remarkable consequence of the interplay of electron correlations, hopping, and orbital energy offset is the fact that {\it{the singlet state is a mixture
of states localized on the same dot with states present on both \cite{rfulde}}}, 
The wavefunction coefficients $\alpha$, $\beta$, and $\delta$ signify 
the contribution of each configuration and can be varied by tuning the DQD's internal structure parameters.}
\label{fig_5}
\end{figure}

{\it{Case I: Coulomb Staircase:}} In first scenario shown in fig.7a,
$\Delta \epsilon = 0$, $U_{11}=U_{22}$, which implies $\kappa=\xi=\frac{1}{\sqrt{2}}$ and $\beta=\delta$. A straightforward evaluation of
coherence factors from Eqs~\ref{eq:rate_bs},~\ref{eq:rate_bt} implies that the criterion for NDR given in
Eq.~\ref{eq:sbc1},~\ref{eq:sbc2} is never satisfied, leading to regular
CB staircase I-V characteristics shown in fig.8a \cite{rbhasko2}. In this case, both left and right 
contacts act are equally efficient in electron-addition and removal processes and hence no such blocking states 
can be formed. Thus the NDR phenomenon is absent.

{\it{Case II: Forward bias Current Collapse:}} In this case $2 \epsilon_2 +U_{22} \approx \epsilon_1 + \epsilon_2 +U_{12}$, which implies that the two electron state $\mid b \bar{b} \rangle$ is almost degenerate with $(\frac{1}{\sqrt{2}}( \mid a \bar{b} \rangle - \mid \bar{a} b \rangle )$ implying that both electrons, {\it{whenever in a singlet state}} can either reside inside dot 2 alone or  one in both dots. The condition $\epsilon_1 > \epsilon_2$ makes it energetically unfavorable to access $\mid a \bar{a} \rangle$ resulting in $\alpha \approx \delta >> \beta$.
Under these conditions, note that our forward bias NDR condition (Eq.~\ref{eq:sbc1},~\ref{eq:sbc1b}) results in
\begin{equation}
\frac{1}{2 \xi^2} > \frac{1}{ {(\kappa \alpha)}^2} + \frac{1}{2 {(\kappa \delta)}^2}
\label{eq:fb}
\end{equation}
which is trivially satisfied in our case $\epsilon_1 > \epsilon_2$. The I-V characteristics now show a
prominent NDR as shown in fig.8b, noted in the experimental trace shown in fig.8d. 
The parameter $\xi$ dictates how effective the right contact 
(also see Eq.~\ref{eq:rate_bt}) is, in the electron removal process between the triplet and bonding states. 
Recall that $\xi << \kappa$ when $\epsilon_1>\epsilon_2$, implying that the right contact has a very 
in-efficient triplet removal, thus blocking the triplet within the DQD system. This causes a current collapse once 
the population of triplet state is energetically feasible. The Forward bias I-V shown in fig.8b shows a strong NDR with current collapse
leading to a small leakage current, once the bias permits the access of transport channel $\epsilon_{10}$ in the bias window.
In the Fock space picture (fig.4c,5), current at lower bias due to $\mid B \rangle \leftrightarrow \mid S \rangle$ drops once the transition $\mid B \rangle \leftrightarrow \mid T \rangle$ occurs since $\mid T \rangle$ is a blocking state. Note that, when transitions $\mid B \rangle \leftrightarrow \mid S \rangle$ and $\mid B \rangle \leftrightarrow \mid T \rangle$ are simultaneously accessed, the system is put in a blocking state once threshold is reached, thereby no NDR occurs. This is shown dotted in fig.8b and occurs experimentally \cite{tar2}, upon application of gate potential, thereby changing the relative position of threshold voltage. We also find that, a finite but small leakage current in the order of $5-10$ pA
occurs due to a finite $\gamma^L_{10} \propto {\xi}^2$ corresponding to the leakage of triplet into a bonding state. Importantly this leakage current reduces by further detuning the dots, making the bonding state more localized on the second dot, such that $\gamma^L_{10} \propto {\xi}^2 \rightarrow 0$. Since $\Gamma=\frac{I_P(V>0)}{I_P(V<0)}>1$ we see a clear asymmetry between the two bias directions in the I-V characteristics.

\begin{figure}
\centerline{\epsfig{figure=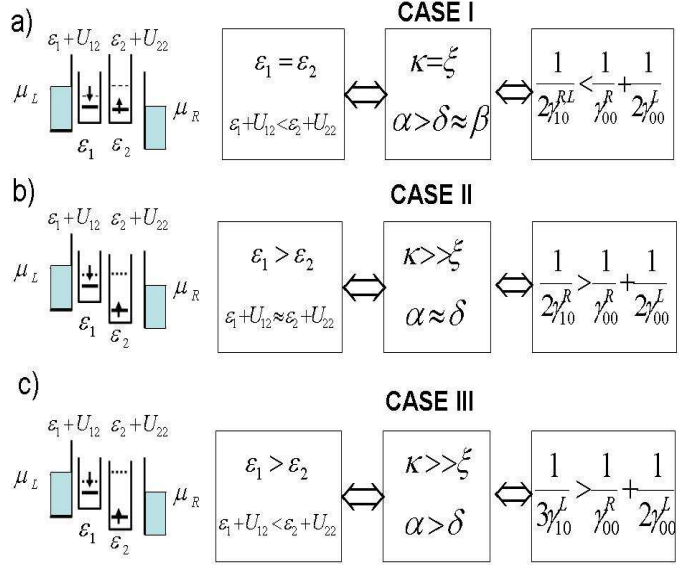,height=3.25in,width=3.6in}}
\caption{Transport scenarios through a DQD system. a) Zero orbital offset $\Delta \epsilon=0$: When $U_{11}=U_{22}>U_{12}$, NDR condition is not satisfied under any bias conditions. b) Finite orbital offset $\Delta \epsilon>0$ with resonance $\epsilon_1 + U_{12} = \epsilon_2 + U_{22}$: Condition for positive bias NDR is satisfied.
c) Finite orbital offset $\Delta \epsilon>0$ and off-resonance
$\epsilon_1 + U_{12} < \epsilon_2 + U_{22}$: the NDR condition is satisfied for both bias directions.}
\label{fig_2}
\end{figure}

{\it{Case III: Multiple Current Collapse:}} Seldom do parameters exactly match the condition $2 \epsilon_2 +U_{22} \approx \epsilon_1 + \epsilon_2 +U_{12}$, experimentally thus making it possible for reverse bias NDR's too. 
This corresponds to the scenario depicted in fig.7c. The important consequence of Eqs.~\ref{eq:NDR},~\ref{eq:NDR2} was that NDR along a bias direction only depends on how the blocking state is coupled to the emptying reservoir.
Now the negative bias ($V_D<0$) NDR condition simply is $\frac{1}{3\gamma^L_{10}} > \frac{1}{\gamma^R_{00}} + \frac{1}{2\gamma^L_{00}}$, which
is also satisfied when $\epsilon_1 + U_{12} < \epsilon_2 + U_{22}$. Under these conditions, shown in fig.7c,8c a mild NDR is noted even in the reverse direction, in good match with experimental (fig.8d,e) observations \cite{tar}.

\begin{figure}
\centerline{\epsfig{figure=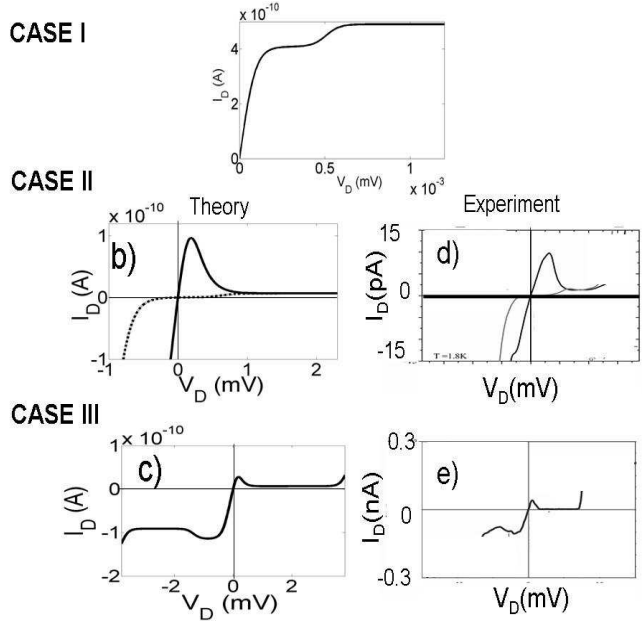,height=3.25in,width=3.6in}}
\caption{I-V Characteristics corresponding to various transport scenarios:
a) Regular Coulomb Blockade plateaus in the absence of orbital off-set, since NDR condition not satisfied.
b) Forward bias NDR occurs when there is finite orbital off-set along with resonance condition corresponding to fig. 7b.
The NDR condition derived in Eq.~\ref{eq:NDR} is physically achieved
in the DQD system by orbital offset such that $\kappa >> \xi$. Furthermore, the resonance
condition $U_{22}-U_{11}=\epsilon_{1}-\epsilon_2$ implies $\alpha=\delta$, thus resulting in
$\frac{1}{\gamma^R_{10}}>>\frac{1}{\gamma^R_{00}} + \frac{1}{\gamma^L_{00}}$.
Also shown dotted is the fact that NDR need not occur when the blocking state is simultaneously accessed with the
conducting state. This can can occur by varying gate voltage \cite{tar2}. c)
Reverse bias NDR's can also occur in the presence of finite orbital off-set with off-resonance discussed in fig.7c.
d) and e) Experimental trnces \cite{tar,tar2} corresponding to scenarios b) and c).}
\label{fig_2}
\end{figure}
 
The experimental I-V's also show a lifting of spin blockade followed by successive NDR's at higher positive bias directions. The lifting of NDR's can be explained simply due to access of transport channels between two electron higher excitations and one electron antibonding states. A consistent match with experimental features, will however involve the inclusion of more orbitals within each dot, such as p-type orbitals, which we shall address in a future work. The fundamentals of NDR's, collapse and rectification still remain within our general framework derived in this paper.
\section{Conclusion}
In this paper, we have developed a general model for multiple current collapse (NDRs)
to occur in the I-V characteristics of strongly interacting systems using the idea of blocking transport channels. With the aid of this model, we have provided an interpretation of
non-linear transport through weakly coupled quantum dots in the Pauli spin blockade regime.
Our model captures subtle experimental observations in this regime, that includes multiple current collapses which are gateable, leakage currents, and rectification, all of which are consistent with the experimental double quantum dot setup. We believe that this model can be exploited in the understanding of other novel blockade mechanisms that can arise from different internal degrees inside a Coulomb Blockaded system. 

Acknowledgments: It is a pleasure to acknowledge Prof. Avik Ghosh for stimulating discussions. This project was supported by 
DARPA-AFOSR.
\section{Appendix I : Derivation of Plateau Currents}
The plateau currents $I_{p1}$ and $I_{p2}$ shown in Fig. 1c are the saturation currents due to
transitions $\mid A \rangle \leftrightarrow \mid B \rangle$, and $\mid A \rangle
\leftrightarrow \mid C \rangle$ whose one-electron energies are $\epsilon_1 =
E_B-E_A$ and $\epsilon_2=E_C - E_A$ respectively. For plateau currents,
(say) under positive bias ($\mu_L > \mu_R$ shown in fig. 1b), additive
transitions rates are given by $R_{A \rightarrow (B,C)}=R^{L}_{A \rightarrow (B,C)}$
and removal transitions are given by $R_{(B,C) \rightarrow A}=R^{R}_{(B,C) \rightarrow
A}$. Furthermore, in order to evaluate $I_{p1}$, given a sequential access of the
two transitions, only states $\mid A \rangle$ and $\mid B \rangle$ appear in
Eq.~\ref{eq:m1}, and ~\ref{eq:curr}. A steady state solution of this master
equation Eq.~\ref{eq:m1}, with $P_C = 0$ along with $P_A + P_B=1$ gives:
\begin{eqnarray}
P_A &=& \frac{R_{B \rightarrow A}}{R_{A \rightarrow B} + R_{B \rightarrow A}} \nonumber\\ 
P_B &=& \frac{R_{A \rightarrow B}}{R_{A \rightarrow B} + R_{B \rightarrow A}},
\label{eq:probs1}
\end{eqnarray} 
Using Eq.~\ref{eq:curr}, and the fact that $f^L(\epsilon_1)=1$ and $f^R(\epsilon_1)=0$ for positive bias, plateau current 
$I_{p1}$ is expressed as:
\begin{eqnarray}
I_{p1} &=& \frac{q^2}{\hbar}\left(R^{L}_{A \rightarrow B} P_A \right) \nonumber\\
\qquad &=& \frac{q^2}{\hbar}\left(R^{R}_{B \rightarrow A} P_B \right) \nonumber\\
\qquad &=& \frac{q^2}{\hbar}\frac{R^{L}_{A \rightarrow B} R_{B \rightarrow A}}{R_{A \rightarrow B} + R_{B \rightarrow A}},
\label{eq:cur1a}
\end{eqnarray}
At a higher bias, the transport channel $\epsilon_2=E_C-E_A$, due to transition $\mid A
\rangle \leftrightarrow \mid C \rangle$ is accessed and thus we have to evaluate the stationary solution of the three state
master equation Eq.~\ref{eq:m1}. 
The steady state solution of Eq.~\ref{eq:m1}, along
with $P_A+P_B+P_C=1$ yields the non-equilibrium probabilities of the three
states as 
\begin{eqnarray}
P_A &=& \frac{1}{1 + \frac{R_{A \rightarrow B}}{R_{B \rightarrow A}} + \frac{R_{A \rightarrow C}}{R_{C \rightarrow A}}}, \nonumber\\ 
P_B &=& \frac{R_{A \rightarrow B}}{R_{B \rightarrow A}} \frac{1}{1 + \frac{R_{A \rightarrow B}}{R_{B \rightarrow A}} + \frac{R_{A \rightarrow C}}{R_{C \rightarrow A}}}, \nonumber\\ 
P_C &=& \frac{R_{A \rightarrow C}}{R_{C \rightarrow A}} \frac{1}{1 + \frac{R_{A \rightarrow B}}{R_{B \rightarrow A}} + \frac{R_{A \rightarrow C}}{R_{C \rightarrow A}}}
\label{eq:probs2} 
\end{eqnarray} 
Similarly as in Eq.~\ref{eq:cur1a}, with the fact that $f^L(\epsilon_2)=1$ and $f^R(\epsilon_2)=0$, 
the second plateau current $I_{p2}$ is expressed as:
\begin{eqnarray}
I_{p2}&=& \frac{q^2}{\hbar}\left(R^L_{A \rightarrow B} P_A + R^L_{A \rightarrow C} P_A \right ) \nonumber\\
\qquad &=& \frac{q^2}{\hbar}\left(R^R_{B \rightarrow A} P_B + R^R_{C \rightarrow A} P_C \right ) \nonumber\\
\qquad &=& \frac{q^2}{\hbar} \frac{R^L_{A \rightarrow B} + R^R_{A \rightarrow C}}{1 + \frac{R^L_{A \rightarrow B}}{R^R_{B \rightarrow A}} + \frac{R^L_{A \rightarrow C}}{R^R_{C \rightarrow A}}}.
\label{eq:cur2a}
\end{eqnarray}
The expressions derived above are cast in terms of rates in general. One has to observe that each plateau current corresponds to a situation when only one contact contributes to the addition of an electron while the other to the removal. 
This implies $f^L(\epsilon)=1$ and $f^R(\epsilon)=0$ for positive bias ($\mu_L>\mu_R$) and vice-versa. Hence the rates, using Eq.~\ref{eq:rates} 
are now given by:
\begin{eqnarray}
R^L_{A \rightarrow B} = \gamma^L  M^L_{AB} =\gamma^L_{AB}, \nonumber\\
R^L_{A \rightarrow C} = \gamma^L  M^L_{AC} =\gamma^L_{AC}, \nonumber\\
R^R_{B \rightarrow A} = \gamma^R  M^R_{AB} =\gamma^R_{BA}, \nonumber\\
R^R_{B \rightarrow C} = \gamma^R  M^R_{AB} =\gamma^L_{CA}, \nonumber\\
\end{eqnarray}
Under these conditions it is easy to see that $I_{p1}$ and $I_{p2}$ follow the expressions in Eq.~\ref{eq:curs}, 
and that conditions for NDR's under positive and negative bias conditions (Eqs.~\ref{eq:NDR}~\ref{eq:NDR2}) 
are easily obtained by replacing $\gamma^L_{ij}$'s by $\gamma^R_{ij}$'s.


\begin{thebibliography}{1}
\bibitem{tar} K. Ono, D. G. Austing, Y. Tokura, and S. Tarucha, Science {\bf{297}}, 1313 (2002).

\bibitem{tar2} K. Ono and S. Tarucha, Phys. Rev. Lett. {\bf{92}}, 256803 (2004).

\bibitem{pet} J. R. Petta, A. C. Johnson, J. M. Taylor, E. A. Laird, A. Yacoby, M. D. Lukin, C. M. Marcus, M. P. Hanson, and A. C. Gossard ,  Science  {\bf{309}},
 2180 (2005).

\bibitem{cm} A. C. Johnson, J. R. Petta, J. M. Taylor, A. Yacoby, M. D. Lukin, C. M. Marcus, M. P. Hanson, and A. C. Gossard, Nature {\bf{435}}, 925 (2005).

\bibitem{kw} F.H.L. Koppens, C. Buizert, K.J. Tielrooij, I.T. Vink, K.C. Nowack, T. Meunier, L.P. Kouwenhoven and L.M.K. Vandersypen, Nature {\bf{442}}, 776 (2006).

\bibitem{het} M. H. Hettler, W. Wenzel, M. R. Wajewijs, and H. Schoeller, Phys. Rev. Lett. {\bf{90}}, 076805 (2003).

\bibitem{dar} E. Vaz and J. Kyriadidis, cond-mat/0608272.

\bibitem{esaki}  L. Esaki Phys. Rev. {\bf{109}}, 603 (1958).

\bibitem{cap} see for example: F. Capasso, K. Mohammad and A. Y. Cho, IEEE J. Quantum Electron. {\bf{QE-22}}, 1853 (1986), and references therein.

\bibitem{hers}N. P. Guisinger, M. E. Greene, R. Basu, A. S. Baluch, and M. C. Hersam, Nano Lett., {\bf{4}}, 55 (2004).

\bibitem{tour} J. Chen, M. A. Reed, A. M. Rawlett. J. M. Tour, Science, {\bf{286}}, 1550 (1999).

\bibitem{kiehl} R. A. Kiehl, J. D. Le, P. Candra, R. C. Hoye, and T. R. Hoye, Appl. Phys. Lett., {\bf{88}}, 172102, (2006).

\bibitem{Heer} H. B. Heersche, Z. de Groot, J. A. Folk and H. S. J. van der Zant, Phys. Rev. Lett., {\bf{96},} 206801, (2006).

\bibitem{rom} C. Romeike, M. R. Wegewijs and H. Schoeller, Phys. Rev. Lett., {\bf{96}} 196805, (2006). 

\bibitem{timm} F. Elste, and C. Timm, Phys. Rev. B {\bf{73}} 235305 (2006).

\bibitem{rralph} E. Bonet, M. M. Deshmukh and D. C. Ralph, Phys. Rev. B {\bf{65}}, 045317
(2002); C. W. J. Beenakker, Phys. Rev. B {\bf{44}}, 1646 (1991), F. Elste and C. Timm {\bf{71}} 155403 (2005).

\bibitem{fran} J. Fransson and M. Rasander Phys. Rev. B, {\bf{73}} 205333 (2006),

\bibitem{plat} J. Inarrea, G. Platero, and A. H. MacDonald, cond-mat/0609323.

\bibitem{plat2} E. Cota, R. Aguado, and G. Platero, Phys. Rev. Lett., {\bf{94}}, 107202, (2005).

\bibitem{sand} I. S. Sandalov, O. Hjortstam, B. Johansson, and O. Eriksson, {\bf{51}}, 13987, (1995). 

\bibitem{bru} H. Bruus and K. Flensberg, Many-body Quantum Theory in Condensed Matter Physics, Oxford University Press, Oxford (2004).

\bibitem{mat} M. Turek, and K. Matveev,  Phys. Rev. B, {\bf{65}}, 115332, (2002).

\bibitem{mitra} A. Mitra, I. Aleiner, and A. J. Millis, Phys. Rev. B, {\bf{69}} 245302, (2004). 

\bibitem{koch} J. Koch, F. von Oppen, Y. Oreg, and E. Sela, Phys. Rev. B {\bf{70}}, 195107, (2004). 

\bibitem{mwl} Y. Meir, N. S. Wingreen, and P. A. Lee, Phys. Rev. Lett {\bf{66}}, 3048 (1991).

\bibitem{bask_the} B. Muralidharan (unpublished).

\bibitem{rbraig} S. Braig and P. W. Brouwer, Phys. Rev. B {\bf{71}}, 195324 (2005).

\bibitem{koenig} see for example: M. Braun, J. K\"onig, and J. Martinek. Phys. Rev. B, {\bf{70}}, 195345, (2004), 
and C. Timm and F. Elste, Phys. Rev. B {\bf{73}}, 235304 (2006) and references therein.


\bibitem{rfulde} `Electron Correlations in Molecules and Solids', P. Fulde,
Springer Series in Solid-State Sciences 100, Springer-Verlag Berlin Heidelberg, 1991.

\bibitem{rbhasko2} B. Muralidharan, A. W. Ghosh, and S. Datta, J. Mol. Simul., {\bf{32}}, 751, (2006).



\end{thebibliography}
\end{document}